\documentclass{aa}
\usepackage{graphics, psfig, pslatex}

\begin{document}

\title{Spin-down of Relativistic Stars with Phase Transitions and PSR J0537-6910}

 \author{N. K. Spyrou \and  N. Stergioulas}

\institute{ Astronomy Department,
Aristoteleion University of Thessaloniki,
541.24 Thessaloniki, Macedonia, Greece \\
 {\it email: spyrou@astro.auth.gr, niksterg@astro.auth.gr}
}

\date{Received / Accepted}

\titlerunning{Spin-down of Relativistic Stars with Phase Transitions}
\maketitle

\maketitle

\begin{abstract}
  
  Using a highly accurate numerical code, we study the spin down of
  rotating relativistic stars, undergoing a quark deconfinement phase
  transition. Such phase transitions have been suggested to yield an
  observable signal in the braking index of spinning-down pulsars,
  which is based on a ``backbending'' behaviour of the moment of
  inertia. We focus on a particular equation of state that has been
  used before to study this behaviour, and find that for the
  population of normal pulsars the moment of inertia does not exhibit
  a backbending behaviour. In contrast, for supramassive millisecond
  pulsars a very strong backbending behaviour is found. Essentially,
  once a quark core appears in a spinning-down supramassive
  millisecond pulsar, the star spins up and continues to do so until
  it reaches the instability to collapse. This strong spin-up
  behaviour makes it easier to distinguish a phase transition in such
  pulsars: a negative first time-derivative of the rotational period,
  $\dot P <0$, suffices and one does not have to measure the braking
  index. In the spin-up era, the usually adopted spin-down power law
  fails to describe the evolution of the angular velocity. We adopt a
  general-relativistic spin-down power law and derive the equations
  that describe the angular velocity and braking index evolution in
  rapidly rotating pulsars.  We apply our numerical results to the
  fast young pulsar J0537-6910 in SNR N157B, which has been suggested
  to have (if spun down by magnetic dipole radiation only) an
  extremely small initial spin period. The inclusion of a quark-hadron
  phase transition can yield a significantly larger initial spin period
  of 6ms (in our example), which is in better agreement with
  theoretical expectations.  Finally, we suggest that the frequent
  rate of glitches in PSR J0537-6910 could be related to the fact that
  it is the fastest Crab-like pulsar, so that a pure quark core may
  have formed recently in its lifetime.

\end{abstract}

\section{Introduction}
\label{introduction}

The properties of the highest-density region of matter in the interior
of relativistic stars have been studied to date by several different
theoretical models, which yield very different macroscopic properties
of compact stars.  Many new interactions have been proposed to occur
at the highest densities attained, ranging from the occurrence of
hyperons, to quark deconfinement, to the formation of kaon condensates
and H-matter (see Weber, 2001; Heiselberg 2002, for recent reviews).
Observational constraints on pulsar properties are currently still too
weak to allow to distinguish which (if any) of all the proposed
theoretical descriptions of the interior of relativistic stars is the
correct one.  A solution to the problem of determining the correct
equation of state in the central region of a compact star must come
from new, more accurate observations of their properties and any new
observational method that will help in this direction is more than
welcome.

One new observational method, that could be interesting in the case
that the equation of state (EOS) features a deconfinement phase
transition to quark matter, has been proposed by Glendenning, Pei and
Weber (1997), (hereafter GPW). Specifically, when a rapidly rotating
pulsar spins down, its central density increases with time. At a
certain central density, a mixed quark-hadron phase can appear (or can
already be present in the pulsar).  For some chosen parameters of the
equation of state and of the baryonic mass of the pulsar, a pure quark
core appears when the pulsar reaches very high densities. In GPW.  a
particular spin-down sequence was studied, that in the nonrotating
limit reached nearly at the maximum mass allowed by the chosen
EOS. For this sequence, it was noticed that, when the pure quark core
appears, the star undergoes an brief era of spin-up (brief, compared
to the pulsar's lifetime) and its braking index shows an anomalous
behaviour (it becomes singular at a certain rotation rate). This
change in the braking index has been proposed in GPW to be potentially
observable.  The anomalous behaviour of the spin evolution is traced
back to a ``backbending'' of the moment of inertia (as a function of
angular velocity).

Other authors (Heiselberg \& Hjorth-Jensen 1998, Chubarian et al.
2000) have further investigated the observational consequences in the
models proposed in GPW and in Glendenning and Weber (2000). In all
these studies the relativistic slow-rotation approximation is
used.\footnote{After this work was completed, we learned of a new
study using rapidly rotating relativistic stars, with a numerical code
related to ours, (Cheng, Yuan \& Zhang, 2002), in which the
backbending behaviour of the moment of inertia is shown to be
sensitive to the assumed properties of the crust.}  In the present
work, we re-investigate the deconfinement phase-transition of
spinning-down pulsars, using fully relativistic, rapidly rotating
models and find that, when the quark-core appears (in a sequence of
models similar to the one used in GPW), the {\it backbending} in the
moment of inertia is generically absent for normal pulsars, but very
strong for supramassive millisecond pulsars. Furthermore, our
numerical results suggest that in the limiting region between normal
and suprammasive pulsars, the backbending behaviour is sensitive to
truncation errors of the applied numerical scheme. The moment of
inertia is a very sensitive function of angular velocity and must be
computed with high accuracy.  Truncation errors, such as a limited
number of points in the EOS table, or the use of the slow-rotation
approximation can be large enough to systematically affect the
computation of the moment of inertia. 

The case of supramassive millisecond pulsars is very interesting, as
the spin-up era lasts from the onset of the quark-core appearence to
their collapse to black holes, i.e. for their remaining lifetime.  For
supramassive millisecond pulsars created via accretion-induced
spin-up, the spin-up era could last for the most part of their
lifetime, if a quark-hadron phase transition occurs. Thus, the {\it first
time-derivative} of the rotational period can be used as a tool for
observing quark-hadron phase transitions and one does not have to rely
on the measurement of the braking index. We emphasize that the
extended spin-up era in supramassive millisecond pulsars is not
restricted for the case of a quark-hadron phase transition, but it can
also occur for other types of phase transitions, such as a phase
transition to a pion condensate. An example is EOS M (Pandharipande
and Smith, 1975) for which an extended spin-up with angular momentum
loss (for supramassive, constant baryonic mass sequences) has been
found in Cook, Shapiro and Teukolsky (1994).  On the other hand, the
absence of spin-up in the currently known millisecond pulsar population
favours the absence of large phase transitions or implies that the
EOS is such that millisecond pulsars created via accretion-induced
spin-up are probably not supramassive. 

In previous studies of spinning-down pulsars that were limited to the
slow-rotation approximation (including up to $O(\Omega^2)$ rotational
effects) expressions for the evolution of angular velocity and braking
index were derived (see Glendenning, 1997). However, we find that
these expressions are {\it incomplete}, as changes in the gravitational mass
of the star (other than changes in the kinetic energy) were not taken
into account in the assumed spin-down law.  Here, we adopt a
general-relativistic version of the spin-down law, which dictates that
the energy lost in the form of e.g. magnetic dipole radiation or
gravitational waves is not lost to the expense of the kinetic energy
only, but to the expense of the total mass-energy of the star. Hence,
making use of a relation due to Bardeen (1970), which expresses the
first law of thermodynamics along sequences of constant baryonic mass,
we derive expressions for the angular velocity and braking index
evolution, which are valid for any rotation rate. In addition, our
adopted spin-down law is also valid during a spin-up era, while 
it's Newtonian, slow-rotation limiting expression (which is usually
adopted in the literature) is not valid during spin-up.

As an application of our numerical investigations, we compute a
possible value for the initial period of pulsar J0537-6910, assuming
that the high-density EOS exhibits a quark-deconfinement phase
transition and that the star spins down only due to magnetic dipole
radiation.  Without the assumption of a phase transition, previous
studies suggested that the initial spin of the pulsar would be
extremely small (for a braking index of $n=3$) (Marshall et al. 1998).
Here we show that the presence of the phase transition can yield a
more reasonable initial period for this pulsar, without having to
assume a braking index smaller than $n=3$.  In our example, we obtain
a significantly larger initial period of 6ms, which is in better
agreement with theoretical expectations. Furthermore, this pulsar has
displayed a large number of giant glitches (Gotthelf et al. 2001). We
propose that these could be due to the appearance of a pure quark core
in the center of the star, which results in an increased rate of
contraction of the star during its spin-down. In contrast, rapidly
rotating pulsars do not exhibit such giant glitches, when their
central density is less than the critical density for the appearance
of the quark core. The fact that pulsar J0537-6910 displays such
glitches more frequently than other Crab-like\footnote{In this paper,
``Crab-like'' only refers to the young age of a pulsar and to its
observational association with a supernova remnant.} pulsars may be
related to the fact that it is the most rapidly spinning of all
Crab-like pulsars and thus a pure quark core could have appeared more
recently in its lifetime.

The outline of the paper is as follows: In Sect. \ref{EOS}, we
describe the chosen equation of state and its numerical refinement,
while in Sec.  \ref{Sequence} we obtain the equilibrium properties of
a specific sequence of spinning-down compact stars. The behaviour of
the moment of inertia and gravitational mass along the sequence is
described in Sec.  \ref{Inertia} and thermodynamic
consistency is used as an independent test in Sec.  \ref{s:Bardeen}. In
Sec. \ref{Braking} expressions for the evolution of the angular
velocity and the braking index are derived, and in Sec. \ref{Initial}
the initial period of PSR J0537-6910 is estimated. Finally, in Sec.
\ref{EventRate}, we comment on the expected event rate of the
considered signal. We conclude with a discussion of our results in
Sec. \ref{Discussion}.

%
\section{Equation of State and Numerical Refinement}
%
\label{EOS}

The EOS used in GPW features a mixed quark-hadron phase, in between
pure hadron and pure quark phases. The mixed phase is obtained by
applying the Gibbs criteria for two-component systems, abandoning
local (but not global) charge-neutrality (Glendenning 1992).  This
yields a second-order phase transition between each region. However,
it has been suggested (see Heiselberg \& Hjorth-Jensen, 1999) that the
mixed phase can exist only if the interface tension between quarks and
hadrons is not too large. But, on the other hand, the interface
tension may not be a free parameter, but it could have the freedom to
adjust (via a minimization prinicple) to the nature of the two phases
(Glendenning 2002), see Christiansen, Glendenning \& Schaffner-Bielich
(2000) for a relevant example.

\begin{figure}[htb]
\centerline{\psfig{file=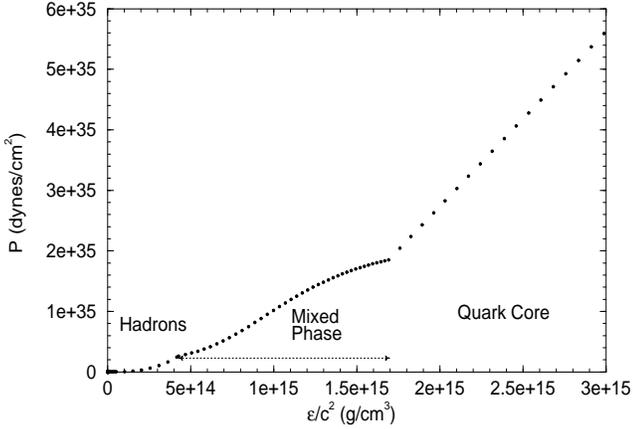,height=2.3in,width=3.3in}}
\caption{Tabulated points of pressure $P$ vs. energy density $\epsilon$ 
for the quark deconfinement EOS. }
\label{eos}
\end{figure}

The specific EOS we use is the quark deconfinement EOS given in Table
9.2 of Glendenning (1997) (same as in GPW) in which, however, we
replace the pure quark phase by an analytic expression (see discussion
below). The pure confined hadronic phase exists below baryon number
density $0.246 {\rm fm}^{-3}$ while pure deconfined quark matter
exists above $0.862 {\rm fm}^{-3}$ (1fm=$10^{-13}$cm). In between
these two phases, a mixed quark-hadron phase exists (see Fig.
\ref{eos}, which displays the pressure $P$ vs. energy density $\epsilon$
relation for this EOS).

\begin{figure}[htb]
\centerline{\psfig{file=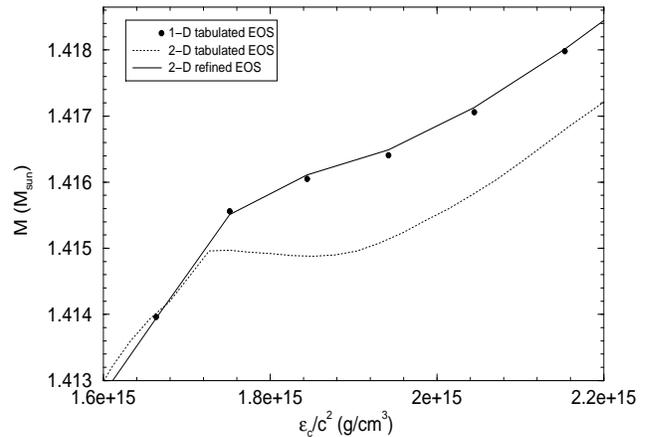,height=2.3in,width=3.3in}}
\caption{Gravitational Mass $M$ vs. central energy density $\epsilon_c$ for 
  nonrotating models, computed by various methods: using the original
  tabulated EOS in the 2-D code produces an oscillatory behaviour
  (dotted line). In contrast, using the refined tabulated/analytic EOS in
  the 2-D code (solid line) gives results in agreement with the
  original tabulated EOS in the 1-D code (when interpolation
  through the phase transition is avoided) (filled circles).}
\label{nonrot}
\end{figure}

When computing nonrotating models with the original tabulated EOS,
using the 2-D code by Stergioulas \& Friedman (1995), we noticed that
for central energy densities near the central energy density where the
pure quark phase appears, the gravitational mass vs. central energy
density plot shows an unusual behaviour (see Fig.
\ref{nonrot}). There is a local maximum followed by a local minimum
and the mass increases again, before it reaches the maximum mass
allowed by the general-relativistic radial instability (see Shapiro \&
Teukolsky, 1983).  Such a behaviour is not present, for exactly the
same tabulated EOS, when using a simple 1-D code that integrates the
well-known radial TOV equations of structure for non-rotating
models. Thus, the oscillatory behaviour of the results obtained with
the 2-D code for nonrotating models is a numerical effect. In the 2-D
code, the angular behaviour of all variables is expanded as a sum of
Legendre polynomials $P_l(\mu)$ of argument $\mu =\cos \theta$.  At
discontinuities, such as the surface of the star (where e.g. the
velocity profile is discontinuous), the accuracy of the code is
reduced due to the {\it Gibbs phenomenon} (see Nozawa et al. 1998).
Similarly, when the pure quark phase appears at the center of the
star, discontinuities appear in the numerical solution, as the EOS
features an abrupt change of slope (see Fig. \ref{eos}). In addition,
as the EOS table only contains a limited number of points, the
interpolation methods (such as 4-point Lagrange interpolation or
Hermite polynomial interpolation) that are used in our code, do not
avoid an oscillatory behaviour in the interpolated quantities, near
the discontinuity in the slope of the EOS. It appears that, in the 2-D
code, such an interpolation error enhances the Gibbs phenomenon at the
center of the star, when the pure quark core appears, resulting in the
oscillatory behaviour in Fig.
\ref{nonrot}.

To solve this numerical problem, we replace the part of the EOS that represents
the pure quark phase by an analytic expression. In fact, we find that this
part can be represented rather accurately by the following linear relation:
\begin{eqnarray}
P [{\rm dynes/cm}^2] &=& 1.85240 \times 10^{35} \\ \nonumber
 && + 2.885209 \times 10^{20} (\epsilon/c^2 - 1.69100\times 10^{15})
\label{Pan}
\end{eqnarray}
where $P$ is pressure, $\epsilon$ is energy density and $c$ is the speed of
light (in CGS units). In this way, the tabulated EOS is used only in
the region below the pure quark phase and no interpolation is done
through the phase transition, thus eliminating the problem of the
oscillatory behaviour.  For nonrotating models, this refined
tabulated/analytic EOS now reproduces the same mass vs. central energy
density relation with the 2-D code, as the original tabulated EOS with
the 1-D code (see Fig. \ref{nonrot}). Notice that for this comparison
we avoid interpolating through the region where the pure quark
appears, by choosing points either before or after the phase
transition only. This increases the accuracy of computed models (as
compared to interpolating through all points in the table)
even with the 1-D code. 

Thus, we have verified that replacing the pure quark region of the
original tabulated EOS by the analytic expression (\ref{Pan})
eliminates numerical problems in the 2-D code, while still
representing the same EOS (with sufficiently high accuracy).

%
\section{Spin-down Sequence}
%
\label{Sequence}

We study several evolutionary sequences of pulsars, that differ from
each other in the assumed constant baryonic mass. For a baryonic mass
of $M_0=1.555 M_\odot$ we obtain a sequence that terminates at the
maximum mass nonrotating model allowed by the EOS. This sequence is
the limiting sequence between normal pulsars and supramassive pulsars
and is very similar to the sequence studied in GPW.  We assume that a
compact star with a baryonic mass of $M_0$ is rapidly rotating.  The
compact star subsequently spins down due to e.g.  magnetic braking
and/or gravitational radiation emission, and the sequence ends either
at the nonrotating limit (for normal pulsars), or the pulsar
encounters the axisymmetric instability limit to black hole collapse
(see e.g.  Cook, Shapiro and Teukolsky, 1984, for examples of such
sequences for various realistic EOSs).  The sequences are constructed
using a highly accurate numerical code ({\tt rns}, Stergioulas \&
Friedman, 1995) that computes exact numerical solutions of rapidly
rotating stars in general relativity, without any approximation (apart
from the truncation errors of the numerical scheme). Solutions are
obtained on a 2-D numerical grid, with typical grid-size of 800x400 or
1200x400 (radial $\times$angular) points.  Each model is defined by its
central energy density and polar to equatorial axis ratio. Apart from
the requirement of high accuracy for each individual model, essential
to the construction of a constant-baryon number sequence, is the
accuracy with which a model of a given baryonic mass is located.  To
this end, we construct various models by keeping the central energy
density fixed, varying only the axes ratio and use a robust
root-finding method to locate a model of given baryonic mass within a
specified accuracy.  The root-finding method of our choice is Ridders'
method (Ridders, 1979), since the procedure is guaranteed to stay
within the initial brackets (avoiding unphysical parameters) and since
it has a relatively high order of convergence ($\sqrt 2$).  In
practice, about 10-12 individual models are required for the
root-finding method to converge to a model of given baryonic mass,
with a relative accuracy of $10^{-9}$ (the accuracy in computing the
baryonic mass for an individual model is of the order of $10^{-5}$,
but this does not limit the root-finding method to converge to a
specified value with higher relative accuracy).

%
\section{Moment of Inertia and Gravitational Mass}
%
\label{Inertia}

\begin{figure}[htb]
\centerline{\psfig{file=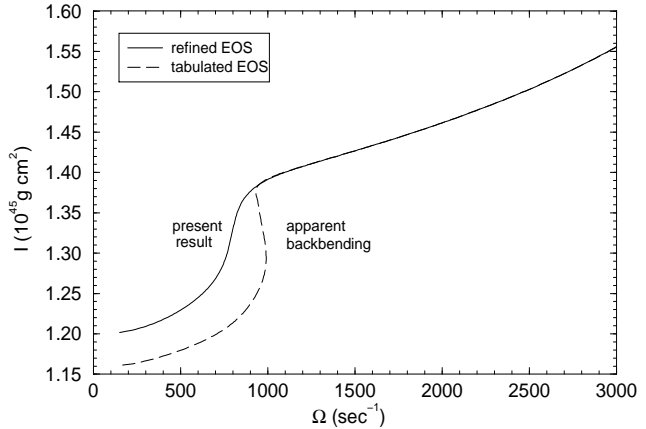,height=2.3in,width=3.3in}}
\caption{Moment of Inertia $I$ vs. angular velocity $\Omega$, along the normal
spin-down sequence with $M_0=1.551 M_\odot$. The original tabulated EOS
produces an apparent backbending in the moment of inertia (dashed
line) while the refined tabulated/analytic EOS yields the evolution of
the moment of inertia without a backbending behaviour.}
\label{f:moment}
\end{figure}

We compute several physical parameters of the equilibrium models, the
moment of inertia $I$, gravitational mass $M$, angular momentum $J$
and angular velocity $\Omega$ being the most relevant for our present
study.  We find that the behaviour of the moment of inertia depends on
the baryonic mass of the sequence. Along a normal sequence with
$M_0=1.551 M_\odot$, which is very close to the maximum mass nonrotating
limit, but still below it, the change in the moment of inertia, as the
star spins down, is shown in Fig. \ref{f:moment}, which plots $I$ as a
function of $\Omega$ (solid line). Using our refined tabulated/analytic
EOS, described above, the behaviour of the moment of inertia is seen
to be regular, i.e. $I$ is a single-valued function of $\Omega$. We notice
that an apparent backbending appears with our 2-D code for this
sequence, when the EOS in the original tabulated form (and
interpolation through all points in the table) is used (dashed line in
Fig. \ref{f:moment}). This is an indication that the backbending of
the moment of inertia is sensitive to the available points in the EOS
table and/or to truncation of rotational effects on the structure of
the relativistic star. 

\begin{figure}[htb]
\centerline{\psfig{file=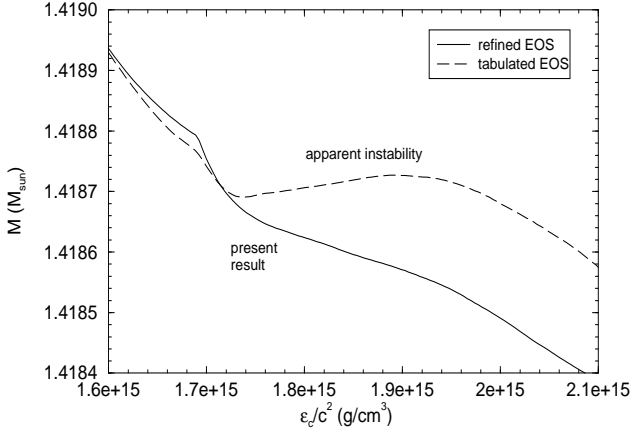,height=2.3in,width=3.3in}}
\caption{Gravitational Mass $M$ vs. central energy density $\epsilon_c$ along the 
normal spin-down sequence with $M_0=1.551 M_\odot$.  Using the original
tabulated EOS produces two turning points (that would correspond to an
unstable region) (dashed line) while the refined tabulated/analytic
EOS produces a monotonous behaviour.}
\label{f:M_Omega}
\end{figure}

Another important equilibrium property that we monitor along the
equilibrium sequence is the gravitational mass. In Fig.
\ref{f:M_Omega}, the gravitational mass $M$ is plotted as a function
of central energy density $\epsilon_c$. The solid line corresponds to the refined
tabulated/analytic EOS and shows a monotonous decrease of the
gravitational mass during spin-down. This normal behaviour corresponds
to the monotonous behaviour of the moment of inertia in Fig.
\ref{f:moment}.  In contrast, when one uses the original tabulated
form of the EOS, the gravitational mass appears to oscillate as a
function of angular velocity.  This oscillatory behaviour is not
correct, as it is the result of numerical truncation errors, due to
the interpolation through the phase transition in the original
tabulated EOS.

The oscillatory behaviour obtained with the original tabulated EOS
deserves some further attention. As we will show, {\it if it were the
  correct behaviour, it would correspond to a part of a sequence being
  unstable to quasi-radial perturbations}: In this sequence, before
the quark core appears, the equilibrium models are stable, while the
star loses kinetic energy $T$ and angular momentum $J$, and the radius
decreases only due to the decrease in rotational flattening of the
star during spin down. When the pure quark core appears at the center
of the star, the radius starts decreasing significantly, as the EOS
becomes softer. Looking at models with larger central densities along
the equilibrium sequence, the angular velocity $\Omega$, angular momentum
$J$, kinetic energy $T$, absolute value of gravitational binding
energy $|W|$ and gravitational mass $M$, all increase. Thus, a star cannot
continue evolving along this uniformly rotating sequence to the
expense of the star's available mass-energy. In Fig. \ref{f:M_Omega}
(dashed lined) a turning point appears and the gravitational mass
increases with increasing central density.  This is a typical case of
the onset of the instability to the fundamental axisymmetric,
quasi-radial mode of oscillation, in rotating relativistic stars, as
shown by Friedman, Ipser \& Sorkin (1988) (hereafter FIS). The
turning-point criterion for the onset of instability was first proved
in FIS for a constant-angular-momentum sequence. Along such a
sequence, an extremum in the $M$ vs.  $\epsilon_c$ curve separates the
regions of stable and unstable stars, i.e.
\begin{equation}
\left(\frac{\partial M}{\epsilon_c}\right)_J =0 ~~ \Leftrightarrow {\rm onset~ of ~instability},
\end{equation}
for a sequence of uniformly rotating stars obeying a one-parameter
$P=P(\epsilon)$ equation of state. 
Such a criterion was then proved by Cook, Shapiro \& Teukolsky
(1992) to also exist for sequences of constant baryonic mass $M_0$:
\begin{equation}
~\left(\frac{\partial M}{\epsilon_c}\right)_{M_0} =0
~~ \Leftrightarrow ~~ {\rm onset~ of ~instability}.
\end{equation}

The instability to the fundamental quasi-radial mode of oscillation is
first a secular one, i.e. it proceeds on a viscous timescale. It has
been shown, however, that a dynamical instability sets in near the
secular instability point (see Shibata, Baumgarte \& Shapiro 2000). Thus,
if the star would evolve along the dashed curve in Fig.
\ref{f:M_Omega}, reaching the secular instability point, it would
first evolve away from the uniformly-rotating equilibrium sequence, on
a viscous timescale, driven by magnetic braking torques and/or
gravitational radiation reaction and could then also become
dynamically unstable, in which case it would suddenly contract.
Eventually, the star would bounce at higher densities, when reaching
again a stable branch in the equilibrium sequence (after the second
turning point in Fig. \ref{f:M_Omega}).

\begin{figure}[htb]
\centerline{\psfig{file=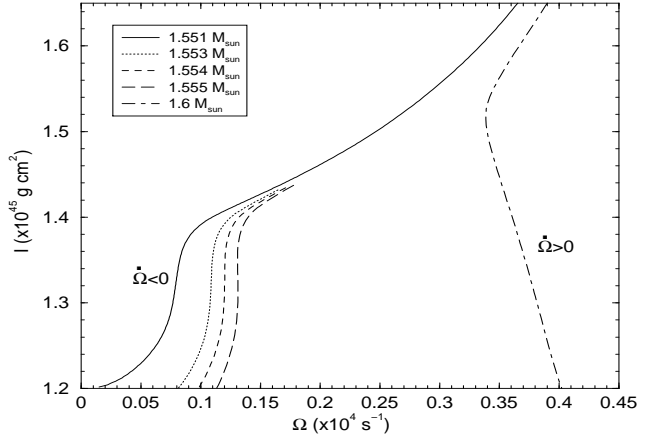,height=2.3in,width=3.3in}}
\caption{Moment of Inertia $I$ vs. angular velocity $\Omega$, along several
spin-down sequences. Normal sequences ($M_0<1.555 M_\odot$) do not show a
backbending behaviour, while for supramassive sequences, such as the
$M_0=1.6 M_\odot$ sequence, the backbending (spin-up) behaviour is dominant.}
\label{f:all}
\end{figure}

\begin{figure}[htb]
\centerline{\psfig{file=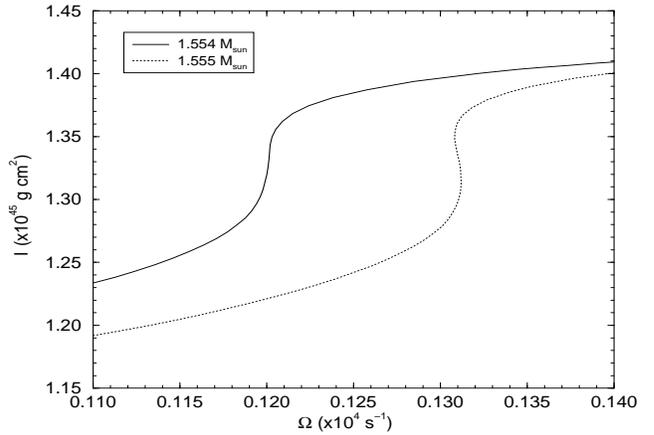,height=2.3in,width=3.3in}}
\caption{Close-up of Figure \ref{f:all}. A small backbending behaviour
first occurs for the limiting sequence of $M_0=1.555 M_\odot$ that
divides normal from supramassive stars and terminates at the maximum-mass
nonrotating model allowed by the given EOS.}
\label{f:all_zoom}
\end{figure}

We find that, when computing sequences of constant baryonic mass, the
behaviour of the gravitational mass can be used as a guide to 
distinguish whether the sequence has been computed accurately: a
behaviour such as the apparent instability described above, indicates
a strong influence of numerical errors in the computation.

Looking at other constant baryonic mass sequences, with baryonic mass
between $M_0=1.551 M_\odot$ and $M_0=1.555 M_\odot$ (see Fig. \ref{f:all}),
we see that the moment of inertia gradually approaches a vertical
behaviour ($\partial I / \partial \Omega \to \infty$), as the limiting sequence that
divides normal from supramassive stars is approached. A very small
backbending (spin-up) behaviour is first observed for the limiting
sequence of $M_0=1.555 M_\odot$ (and clearly not for e.g. $M_0=1.554
M_\odot$, see Fig. \ref{f:all_zoom}). For larger baryonic masses
(supramassive sequences) the backbending behaviour becomes stronger
and for e.g. a baryonic mass of $M_0=1.6 M_\odot$ it becomes the dominant
behaviour, i.e. once the quark core appears, the star only spins-up
with angular momentum loss, for the rest of its lifetime.

To summarize, for the particular EOS that we have studied in this
paper, normal pulsars do not feature a spin-up era during phase
transition, while supramassive pulsars only spin up after the phase
transition. The limiting sequence that divides normal from supramassive
pulsars shows a very small era of spin-up and appears to be the dividing
line between stars that can spin up and stars that only spin down.

%
\section{Thermodynamic Consistency}
%
\label{s:Bardeen}

In the previous sections we demonstrated that  the refined
tabulated/analytic EOS produces a constant baryonic mass sequence with
equilibrium properties that are in sharp contrast to the equilibrium
properties computed on the basis of the original tabulated EOS. In
this section we will use an independent check to verify that the
equilibrium properties obtained with the refined EOS are indeed the
physically acceptable ones, while the equilibrium sequence obtained
with the tabulated EOS suffers from numerical errors. 

The independent check of the accuracy of the computed equilibrium
properties is provided by the use of a relation that was first proved
by Ostriker and Gunn (1969) in Newtonian theory and then derived in
general relativity by Bardeen (1970).

Along a sequence of uniformly rotating models of constant baryonic
mass, changes in $M$ and $J$ are related by
\begin{equation}
c^2dM = \Omega dJ,
\label{Bardeen}
\end{equation}
which can be regarded as an expression of the first law of thermodynamics
for such sequences.

\begin{figure}[htb]
\centerline{\psfig{file=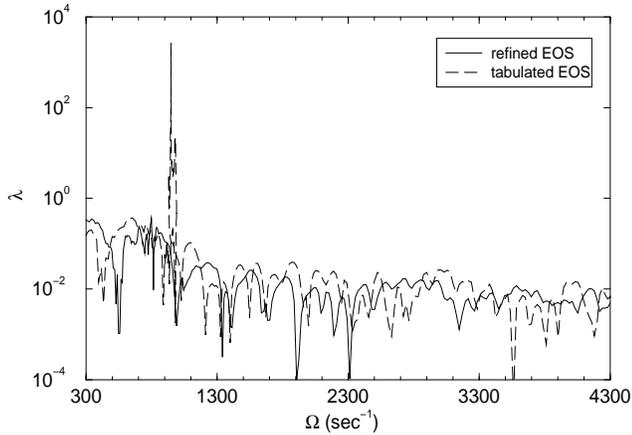,height=2.3in,width=3.3in}}
\caption{Error indicator $\lambda$, constructed with the use of the 
thermodynamic consistency relation (see text) for the original and for
the refined EOSs, along the sequence of rotating models undergoing the
quark-hadron phase transition, with $M_0=1.551 M_\odot$. The refined EOS
gives significantly more accurate models in the region where the pure
quark core appears.}
\label{lambda}
\end{figure}

We have evaluated the above relation numerically along the obtained
sequences of equilibrium models (using sufficiently many individual
models to ensure an accurate evaluation of the numerical derivative).
We define an error indicator, $\lambda$, as
\begin{equation}
\lambda=\left |1-\frac{c^2dM}{\Omega dJ} \right |.
\end{equation}
The evaluation of $\lambda$ along the evolutionary sequence of $M_0=1.551
M_\odot$ (Fig. \ref{lambda}), shows that the above relation is satisfied
with an accuracy of $\sim 1$\% for models with central energy densities
smaller than the energy density at which the pure quark-core
appears. After the appearance of the pure quark core, numerical errors
result in huge values of $\lambda$, when using the original tabulated
EOS. With the use of the refined tabulated/analytic EOS, these errors
are reduced significantly to an acceptable level and diminish as the
resolution is increased.

The above check confirms that the refined EOS indeed produces physically
acceptable results, while the original tabulated EOS produces results
that are dominated by numerical error in the region in question.

%
\section{Behaviour of the Observable Braking Index}
%
\label{Braking}

If a compact star is detected as a pulsar and one can measure the
first two time-derivatives $\dot \Omega$, $ \ddot \Omega$ of its angular velocity,
then one can define an observational braking index 
\begin{equation}
n=n(\Omega)\equiv \frac{\Omega \ddot \Omega}{\dot \Omega^2}. \label{nd}
\end{equation}
In the Newtonian, slow-rotation limit, the spin-down of a pulsar is
usually modeled in the form of a power law (see Shapiro \& Teukolsky.
1983), namely assuming that only kinetic energy is lost and that the
rate of loss is proportional to some power of the angular velocity of
the star, i.e.
\begin{equation}
\frac{dT}{dt} = \kappa \Omega^\alpha,
\label{dT1}
\end{equation}
where $\kappa<0$ and $\alpha>0$ are usually assumed to be real constants. With
these assumptions, the braking index is equal to $n=\alpha-1$.  For
example, it is assumed that, for magnetic braking, $\dot T \sim \Omega^4$,
which yields an expected braking index $n=3$, while, for gravitational
wave emission, $\dot T \sim \Omega^6$, which yields $n=5$. For slowly
rotating pulsars, the above considerations are, in most cases,
appropriate.  However, for rapidly rotating pulsars one has to take
into account the rotational flattening of the star.  To a first
approximation, this can be done considering rotational effects up to
order $O(\Omega^2)$.  Glendenning (1997) gives the rotationally corrected
$\dot \Omega$ and braking index $n(\Omega)$ for a sequence of uniformly
rotating stars, assuming the spin-down law of Equation (\ref{dT1}).
As we will show here, the expressions given in Glendenning (1997) are
incomplete, in the sense that the derivation is not fully consistent
to $O(\Omega^2)$, but misses additional contributions of the same order.

The energy lost in the form of electromagnetic or gravitation
radiation is not only to the expense of the star's kinetic energy
(which would be the case only in the $O(\Omega)$ slow-rotation
approximation) but to the expense of the star's total mass-energy
(gravitational mass).  In the Newtonian limit, this principle has been
applied to the spin-down of rapidly rotating neutron stars by Finn \&
Shapiro (1990) and by Spyrou \& Stergioulas (2001). In relativistic
stars, the total mass-energy is
\begin{equation}
Mc^2 = M_0c^2+U+T+W,
\end{equation}
(see Friedman, Ipser \& Parker 1986) where, additionally, $U$ is the
internal energy.  We thus adopt the following natural generalization of
Equation (\ref{dT1}) in general relativity:
\begin{equation}
c^2\frac{dM}{dt} = \kappa \Omega^\alpha.
\end{equation}

The kinetic energy in relativity is defined as $T=(1/2)J\Omega$, so
that
\begin{equation}
\frac{dT}{dt} = \frac{1}{2} \left( c^2 \frac{dM}{dt} 
               +\frac{d\Omega}{dt}J \right).
\end{equation}
Expressing this in terms of the moment of inertia, which is defined
as in Friedman et al. (1986), $I=J/ \Omega $, and with the help of
Equation (\ref{Bardeen}), we find
\begin{equation}
\frac{dT}{dt} = \left( c^2\frac{dM}{dt} 
               -\frac{1}{2} \Omega^2 \frac{dI}{dt}J \right). \label{dT}
\end{equation}
Equation (\ref{dT}) shows that, in a rapidly rotating star, under the
above assumptions, the time derivatives of the kinetic energy and
gravitational mass differ by an $O(\Omega^2)$ term that is proportional to
the time-derivative of the moment of inertia. Notice that this
expression is exactly valid for any rotation rate (even for stars
rotating at the mass-shedding limit) and not only in an $O(\Omega^2)$
approximation. The relations derived in Glendenning (1997) for the
braking index and time-derivative of the angular velocity in the
$O(\Omega^2)$ approximation, are missing the
contribution from the $O(\Omega^2)$ term on the r.h.s. of Equation
(\ref{dT}).

\begin{figure}[htb]
\centerline{\psfig{file=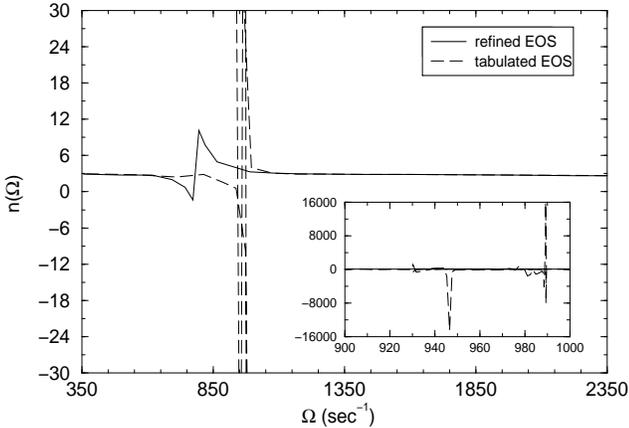,height=2.3in,width=3.3in}}
\caption{Observational braking index $n(\Omega)$ along the spin-down sequence
with $M_0=1.551 M_\odot$. The original tabulated EOS yields a singular
behaviour at two nearby values of the angular velocity (see
inset). The refined EOS does not yield a singular behaviour, but only
a much weaker increase and corresponding decrease (with respect to the
canonical value of $n=3$) at the pure quark core appearance.}
\label{f_n}
\end{figure}

With the above assumptions, it is easy to show that Equations (5.18) and
(5.20) in Glendenning (1997) are replaced by 
\begin{equation}
\dot \Omega = \frac{\kappa}{I} \Omega^{\alpha -1}\left(1+\frac{\Omega I'}{I} \right)^{-1},
\label{Omdot}
\end{equation}
where $I'=dI/d\Omega$ and
\begin{equation}
n(\Omega) = \alpha -1 ~-~ \frac{2I' \Omega + I'' \Omega^2}{I+I'\Omega}. \label{n}
\label{nOmega}
\end{equation}
We have verified that Equation (\ref{nOmega}) is the correct one, by
numerically computing the braking index as obtained above and
comparing it to the direct computation of the braking index, using the
definition Equation (\ref{nd}). The two numerical results agree well
even close to the mass-shedding limit. In contrast, the expression for
$n(\Omega)$ in Glendenning (1997) shows a large disagreement with the
result obtained from the definition of $n(\Omega)$ for rapidly rotating
stars. This difference between the two expressions is significant
also in the case of supramassive sequences.

In Fig. \ref{f_n}, the observational braking index $n(\Omega)$ is shown as
a function of angular velocity (computed along the normal spin-down
sequence with $M0=1.551 M_\odot$), assuming that the spin-down is only
due to magnetic braking.  The computed $n(\Omega)$ is roughly $n\simeq 3$
before the pure quark core appearance. During the phase transition,
the braking index rises to roughly $n \simeq 10$ and then decreases to a
value of roughly $n\simeq -2$, before returning to the canonical value of
$n\simeq 3$. This behaviour is qualitatively and quantitatively different
from the singular behaviour in the braking index obtained with our
code when one uses the original tabulated EOS (see inset in
Fig. \ref{f_n}). 

\section{Initial Period of PSR J0537-6910}
\label{Initial}                                     

PSR J0537-6910 was discovered in SNR N157B by Marshall et al.
(1998).  It has a spin period of only 16ms, the fastest of all
Crab-like pulsars, and a period derivative of $\dot P=5.13 \times
10^{-14}{\rm s s}^{-1}$. Both the characteristic age $\tau =P/(2\dot P)$
and the age estimate for SNR N157B suggest that the pulsar is 5,000
years old. An upper limit on the age of the pulsar based on
observations of $H\alpha$-emitting gas in N157B (Chu et al. 1992), is
20,000 years.  Marshall et al. show that, if the braking index is
$n=3$ and the pulsar age is 5,000 years, then the pulsar must have
been born with an extremely small (even sub-millisecond) period. Such
small initial periods in pulsars associated with supernova remnants
have not been detected to date.  There are theoretical arguments, based
e.g. on the possible occurrence of gravitational-wave induced
instabilities in rapidly rotating relativistic stars (see e.g.
Stergioulas 1998, Andersson \& Kokkotas 2001 for reviews) that suggest
a pulsar initial period (shortly after the compact star is born) of at
least several ms. How can this theoretical expectation be
compatible with the observed properties of PSR J0537-6910 ? One
possible solution would be that the actual braking index for this
pulsar is closer to $n=2$ (for the upper limit on age of 20,000 years,
a very low braking index of $n\simeq 1$ is required).

Here we show that an initial period of several ms (6 ms in our
example) can be obtained for a braking index of $n=3$, if the
appearance of a pure quark core is considered. We use the same EOS and
the $M0=1.551 M_\odot$ constant baryonic mass sequence mentioned in
previous sections and numerically integrate Equation (\ref{Omdot}). We
find that the age of the pulsar can be expressed in the following
integral form
\begin{equation}
t_{\rm age}= - \frac{1}{\kappa}\int_{\Omega_i}^{\Omega}\frac{I+\Omega dI/d\Omega}{\Omega^{\alpha-1}}d\Omega,
\end{equation}
where $\Omega_i$ is the initial angular velocity, or, equivalently,
\begin{equation}
t_{\rm age}= - \frac{1}{\kappa}\int_{J_i}^{J}\Omega^{1-\alpha}dJ,
\end{equation}
where $J_i$ is the initial angular momentum. For numerical integration, it
is more convenient to introduce a dimensionless central energy density
$\tilde \epsilon_c=(\epsilon_c/c^2)/(10^{15}{\rm g/cm^3})$ as the integration parameter, and
evaluate
\begin{equation}
t_{\rm age}= - \frac{1}{\kappa}\int_{\tilde \epsilon_c(i)}^{\tilde \epsilon_c}\Omega^{1-\alpha}\frac{dJ}
{d \tilde \epsilon_c} d \tilde \epsilon_c.
\end{equation}

\begin{figure}[htb]
  \centerline{\psfig{file=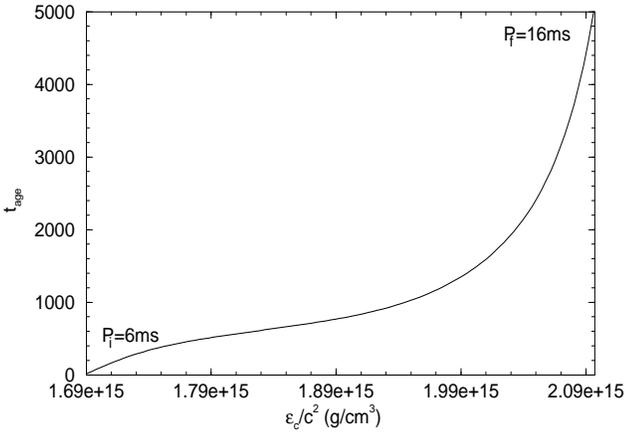,height=2.3in,width=3.3in}} 
\caption{Computed age of PSR J0537-6910 as a function of its central
energy density. For a current age of 5,000 years, the initial spin 
period is 6ms, significantly larger than in the absence of a phase
transition.}
\label{f_age}
\end{figure}

The result of the numerical integration is shown in Fig. \ref{f_age},
which displays the age of the pulsar as a function of the central
energy density. For a current spin of 16ms and a pulsar age of 5,000
years, the initial spin is obtained to be 6ms.  Thus, even an $n=3$
braking index can yield an initial spin that is in agreement with
theoretical expectations, provided the occurrence of a pure quark core
is taken into account. In the above example, a particular choice of
EOS and baryonic mass was made. Different choices would lead to
different initial spin period estimates. However, the qualitative
effect of the presence of a pure quark core is to increase (with
respect to an EOS without a phase transition) the computed initial
spin period.

\section{Event Rate}
\label{EventRate}                                              

An important parameter for the successful detection of a
phase-transition signal in the braking index of spinning-down pulsars
is the event rate, which is determined by several important factors.
Let's assume that pulsars are born with an initial period of only a
few milliseconds. Since normal pulsars have a magnetic field strength
of $\sim 10^{12}G$, they would quickly spin down to much larger periods
due to magnetic dipole radiation. Thus, if the baryonic mass of a
newly-born pulsar is such that the phase transition appears at a short
rotational period, then the phase transition signal will not last for
an extended period of time to have a realistic chance of being
detected. Notice that no young, Crab-like pulsar with period shorter
than $16$ms has been observed, which is indicative of the very small
event rate that should be expected for such an observation.

\begin{figure}[htb]
  \centerline{\psfig{file=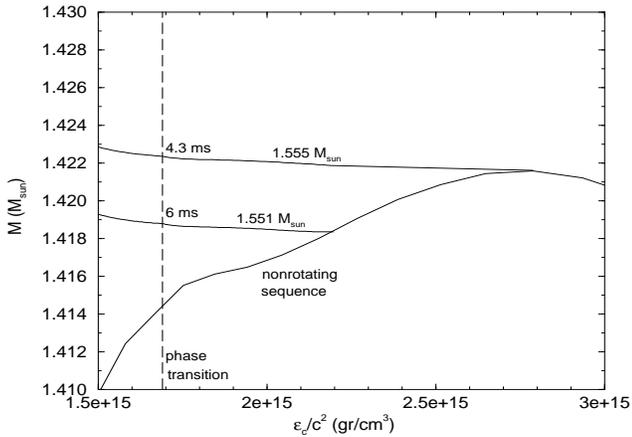,height=2.3in,width=3.3in}} 
\caption{Gravitational mass vs. central energy density $\epsilon_c$ 
for the nonrotating sequence and for the spin-down sequence of
constant baryonic mass $M_0=1.551 M_\odot$. The dashed vertical line
indicates the central energy density at which a pure quark core
appears. Slowly rotating normal pulsars can go through the phase
transition, only if the gravitational mass falls within an extremely
narrow range of values.  }
\label{f_spindown}
\end{figure}

On the other hand, if pulsars are born with rotational periods larger
than, say, 6ms, then a phase transition can occur during their
lifetime only if their baryonic mass falls within an extremely small
range of values.  To illustrate this point, we plot in Figure
\ref{f_spindown} the constant baryonic mass sequence considered in
Sec. \ref{Sequence} and the corresponding sequence of nonrotating
models. The vertical dashed line specifies the central density at
which the pure quark core appears. It crosses the
constant-baryonic-mass sequence at a model with rotational period of
6ms and gravitational mass of $1.419M_\odot$. Notice that the
gravitational mass of a nonrotating star with central energy density
equal to that of the quark core appearance ($1.691\times10^{15}{\rm
  g/cm^3}$) is $1.414M_\odot$. Thus, a pure quark core can appear in a
spinning down pulsar with rotational period less than 6ms only if its
mass falls within $\Delta M\sim 0.0025 M_\odot$ of the value of $1.4165M_\odot$
(the the sequence considered in this example). The probability for
this to happen is obviously negligible.

The above arguments show that the only observed population remaining,
in which one could hope for a significant event rate for the
considered signal, are old millisecond pulsars, spun-up by accretion,
with low magnetic field and spin-down rate. The actual event rate will
then depend on the mass-distribution for this population and on the
precise value of the central energy density at which the quark core
appears. None of these parameters are currently known with sufficient
accuracy to compute a reliable value for the event rate.  In fact, the
observational determination of the braking index is itself a difficult
task and no braking index has been measured for any of the known
millisecond pulsars, to date. However, should a millisecond pulsar
become supramassive, due to mass-accretion, then there is a realistic
chance for detecting the presence of a quark core in observations of
the first time-derivative, $\dot P$, of the rotational period. Once
the quark core appears and if the pulsar is sufficiently massive, 
then $\dot P<0$ for the rest of the pulsar's lifetime.

\section{Discussion}
\label{Discussion}                                              

Observational constraints for the properties of compact stars are
currently still too weak to determine the correct equation of state of
high-density matter.  The presence of a quark/hadron deconfinement
phase transition has been suggested by Glendenning, Pei and Weber
(1997) (GPW) to be observable in the braking index of pulsars. In this
method, the observational signal comes from a  spin-up
era, originating from a backbending behaviour of the moment of inertia
during the pure quark core appearance.

We have investigated the deconfinement phase-transition of
spinning-down pulsars, using fully relativistic, rapidly rotating
models. We find that, for normal pulsars, the star does not go through
a spin-up era and the pure quark core appears without a backbending
behaviour in the moment of inertia.  The backbending behaviour first
occurs for the limiting sequence of constant baryon mass that
terminates at the maximum mass nonrotating model allowed by the
EOS. In contrast, for sequences with larger mass (supramassive
sequences) the backbending behaviour becomes dominant and pulsars
only spin up, after a quark core appears.

From our numerical results, it is evident that the moment of inertia
along a spin-down sequence is very sensitive to the appearance of
phase transitions and must be computed with great care.  In
particular, we emphasize that low resolution and the restriction to a
limited number of points in the EOS table (with interpolation through
all points) can easily spoil the correct computation of the behaviour
of the moment of inertia, at least in the present 2-dimensional code
used in this paper.

The expected event rate of a phase-transition
signal in the pulsar braking index is negligibly small for the
population of normal pulsars. For the population of old millisecond
pulsars, a reliable estimate for the event rate cannot be obtained, as
it involves several uncertain factors and, in addition, the
observational determination of braking indices in millisecond pulsars
has not become technically possible, yet. However, if millisecond
pulsars become supramassive, then the presence of a quark core could
be detected by observing a negative first time-derivative of the
rotational period. On the other hand, the absence of any such observation
in the current population of known millisecond pulsars indicates
that either large phase transitions do not occur, or that the EOS
is sufficiently stiff to prevent accreting relativistic stars from
becoming supramassive. A detailed study of phase transitions in
supramassive models of relativistic stars, constructed with 
various EOSs, could provide interesting constraints on the equation
of state at very high densities.

In order to describe the spin down of a pulsar due to magnetic dipole
or gravitational radiation, we adopt a general-relativistic version of
the usual power-law assumed in the slow-rotation limit. The energy
loss is assumed to be to the expense of the total mass-energy of the
star and not simply to the expense of the star's kinetic energy. In
this way, we derive new expressions for the evolution of the angular
velocity and braking index during spin down. As an application,
assuming magnetic dipole braking only, the initial spin period of PSR
J0537-6910 is obtained to be 6ms (for the EOS and spin-down sequence
considered in this paper). The presence of the deconfinement phase
transition can thus give a more reasonable initial period for this
pulsar, compared to previously obtained estimates.

PSR J0537-6910 exhibits the highest rate of large glitch events (6 in
3 years) (see Gotthelf et al. 2001) of all known Crab-like pulsars.
At the same time, it is the most rapidly rotating Crab-like pulsar. If
a pure quark core appears at small spin periods, then the quark-core
fraction in such pulsars continuously changes during spin down, most
probably contributing to the occurrence of large glitches.  The fact
that PSR J0537-6910 exhibits the highest glitch rate could be related
to the fact that the phase transition may have happened more recently
in its lifetime (according to the spin-down sequence used here) and
its spin-down behaviour is still influenced  by the gradual change of
the quark core fraction during spin down. This suggestion has to be
examined in more detail, as it could be a generic feature of
phase transitions in the high-density EOS.

\section*{Acknowledgements}                                              

We thank Nils Andersson, Pawel Haensel and John L. Friedman for useful
discussions and Emanuele Berti for a careful reading of the
manuscript. We are grateful to Leszek Zdunik and Pawel Haensel for
comparing numerical results of several evolutionary sequences, before
publication. Finally, we are indebted to the referee, Prof. N. Glendenning,
for critical comments that substantially improved the final version
of this paper. This work has been supported by the EU Programme 'Improving
the Human Research Potential and the Socio-Economic Knowledge Base'
(Research Training Network Contract HPRN-CT-2000-00137),
KBN-5P03D01721 and the Greek GSRT Grant EPAN-M.43/2013555.



\end{document}